\documentclass[journal]{IEEEtran}
\ifCLASSINFOpdf
\else
\fi

\usepackage{hyperref}
\usepackage{amsmath}
\usepackage{graphicx,subcaption}

\usepackage{xcolor}

\usepackage[backend=biber]{biblatex}

\begin{document}

\title{Training a Distributed Acoustic Sensing Traffic Monitoring Network With Video Inputs} 

\author{Khen Cohen\IEEEauthorrefmark{1}, Liav Hen\IEEEauthorrefmark{2}, and Ariel Lellouch\IEEEauthorrefmark{3}

\IEEEauthorrefmark{1} khencohen@mail.tau.ac.il; School of Physics and Astronomy, Tel-Aviv University, Tel-Aviv 69978, Israel \\
\IEEEauthorrefmark{2} School of Electrical Engineering, Tel-Aviv University, Tel-Aviv 69978, Israel \\
\IEEEauthorrefmark{3} Porter School of the Environment and Earth Sciences, Tel-Aviv University, Tel-Aviv 69978, Israel
} 


\maketitle

\begin{abstract}
Distributed Acoustic Sensing (DAS) has emerged as a promising tool for real-time traffic monitoring in densely populated areas. In this paper, we present a novel concept that integrates DAS data with co-located visual information. We use YOLO-derived vehicle location and classification from camera inputs as labeled data to train a detection and classification neural network utilizing DAS data only. Our model achieves a performance exceeding $94\%$ for detection and classification, and about $1.2\%$ false alarm rate. We illustrate the model's application in monitoring traffic over a week, yielding statistical insights that could benefit future smart city developments. Our approach highlights the potential of combining fiber-optic sensors with visual information, focusing on practicality and scalability, protecting privacy, and minimizing infrastructure costs. To encourage future research, we share our dataset.
\end{abstract}


\IEEEpeerreviewmaketitle

\section{Introduction}

\begin{figure*}
    \centering
    \includegraphics[width=1.0\linewidth]{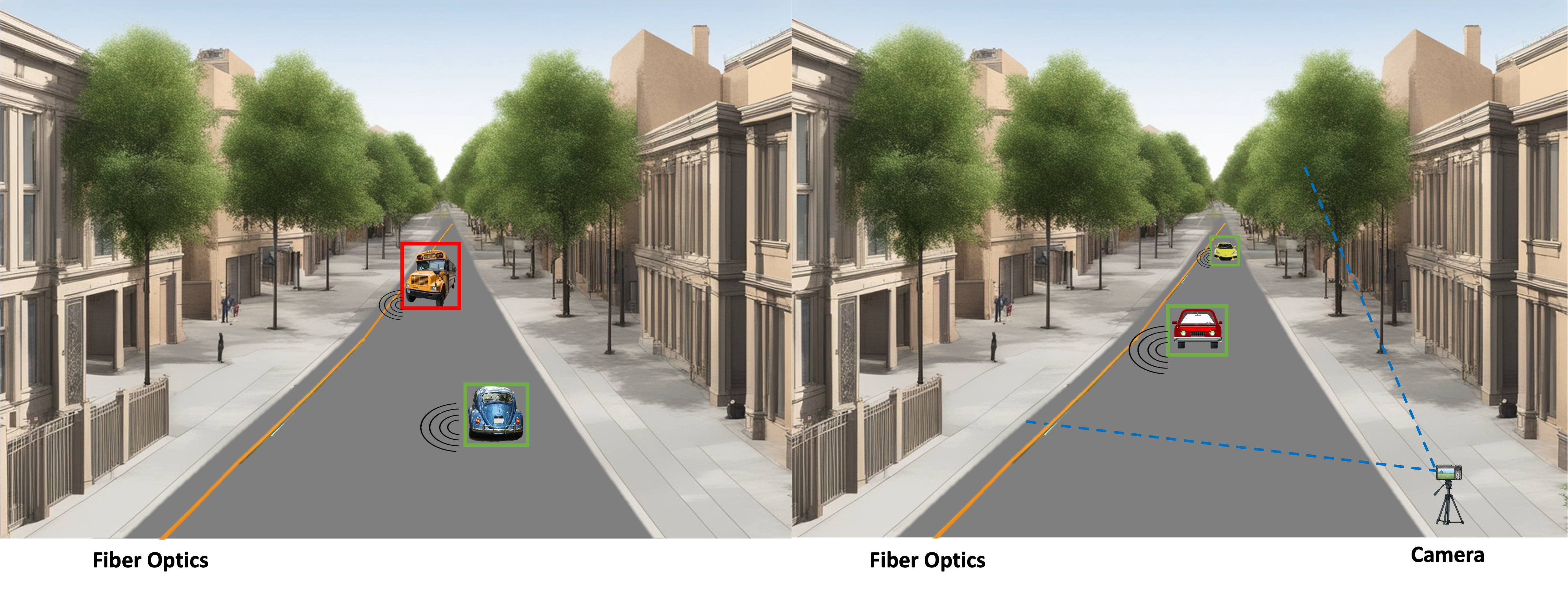}
    \caption{Illustration of proposed approach. Right: the training stage, involving DAS and camera recording simultaneously. The blue rectangles represent the computer vision detection and classification algorithm for each of the objects. Left: the test stage, involving only DAS, in which our algorithm learns to detect, classify, track, and estimate the location of each of the objects. Any moving object which is not part of the defined classes set is considered as noise and filtered out. }
    \label{fig:main_diagram}
\end{figure*}

\IEEEPARstart{D}{istributed} acoustic sensing (DAS) transforms standard telecommunication fiber optic cables into dense seismic arrays. The underlying sensing mechanism relies on Rayleigh backscattering \cite{fosexample}, occurring due to minuscule changes in density and refractive index along the fiber. Through an optical interrogator unit (IU), continuous laser light is pumped into the fiber, and the backscattered light is measured \cite{das4monitoring, introToDfos, fosCommunication}. There are two main approaches to implementing DAS \cite{dasHWtypes, dasHWprogress}. The first is Optical Frequency Domain Reflectometry (OFDR), in which the backscattered pulse is analyzed in the frequency domain, and Optical Time Domain Reflectometry (OTDR), in which it is analyzed in the temporal domain. Among the OTDR methods, a phase OTDR ($\Phi-OTDR$) estimates the phase of the backscattered signal by interferometry. This is a useful property as the phase change of the backscattered light is quasi-linearly proportional to directional strain, or strain-rate, along the direction of fiber. 

The $\Phi-OTDR$ sensing method, which we use in this study, provides very high spatial (in the order of meters) and temporal (in order of KHz) resolution. A single IU can cover distances of tens of kilometers and more \cite{longvalleyarray}. In addition, the instrument response is very broadband, spanning about 17 octaves \cite{dasinstrumentresponse}. Finally, DAS technology allows to re-purpose existing telecommunication infrastructure, significantly reducing efforts and costs of instrument deployment. For these reasons, DAS is well-suited for a wide range of scientific disciplines and practical applications, such as earthquake seismology and early warning systems, hydrocarbon exploration, volcanology and glaciology, geotechnical engineering, structural monitoring, oceanography, security, and more \cite{lindsey2021fiber, das4monitoring, dasantennas, downholedas, das_linear_infra}. 

In addition, the ubiquity of existing fiber optic infrastructure in urban areas and its deployment mainly along roads \cite{daspotential} allow for DAS-based traffic monitoring \cite{trafficflow}. Generally, the imprint of the vehicle weight, whose frequency content is $<$1 Hz, is used for traffic monitoring, as opposed to other DAS applications that utilize wider frequency bands. Roadside DAS can, in general, provide valuable insights into traffic flow, vehicle classification, and road condition assessment \cite{ChiangVehicleDetection1, Kou24VehicleDetection, CoreraVehicleDetection3, traffic-pattern2, traffic-pattern3, traffic-pattern4, roadcondition, vehicle-detection, wang2022}. DAS has several advantages over existing monitoring approaches, which are typically vision-based, include pavement sensing, or rely on mobility (GPS) data \cite{yuan2023,lindsey2020city}. DAS data are intrinsically anonymous and preserve privacy, as opposed to cameras \cite{cameraUrbanTracking, cameraUrbanMonitor1, MARASINGHE2024105047, pedestrianTracking, Camera4Monitoring} or mobility \cite{mobility1,mobility2,mobility3,mobility4} data. They are also complete as they do not depend on weather conditions or mobile phone access and data sharing permissions. The fiber infrastructure is entirely passive, maintained by telecommunications operators, and unlikely to be targeted by vandalism. With the increasing demand for smart city monitoring and efficient transportation management, the long-range sensing potential of DAS, given adequate fiber routing, makes it a cost-effective solution, as opposed to the linearly increasing cost of cameras or pavement sensors.

The majority of traffic monitoring studies using DAS focus on identifying and categorizing various vehicle types, including cars, trucks, and buses, leveraging standard optical fiber infrastructure for traffic monitoring \cite{xie2024intelligenttrafficmonitoringdistributed}. In addition to traditional signal and image processing approaches \cite{chambers2020,lindsey2020city,image-processing-work,objectdetection}, recent progress in the field has seen the implementation of deep learning techniques, particularly convolutional neural networks (CNNs), to analyze DAS patterns and enhance vehicle classification accuracy \cite{yuan2023, vandenende, objectdetection, vehicleextraction, martin2018}. Methods such as U-Net architectures segment DAS waveform data into discernible vehicle paths, providing high-resolution, real-time monitoring across extensive distances. These methods can achieve vehicle identification accuracy rates exceeding $97\%$ under controlled conditions, demonstrating significant potential for high-performance traffic monitoring applications. Nonetheless, challenges persist particularly around data labeling, which is often time-consuming and subjective, complicating the application of fully supervised learning methods. This has led researchers to turn to synthetic dataset generation, simulating DAS signals to supplement labeled data and facilitate model training \cite{yuan2023, shiloh2020}. However, synthetic datasets cannot fully encapsulate the complexity of real data which depends on variable fiber coupling, subsurface geology, imprecise fiber location, noise conditions, fiber geometry, vehicle type and model, and more. Consequently, unsupervised and self-supervised learning approaches have also been explored to address these limitations, though accuracy remains a challenge as these methods struggle with capturing complex noise and diverse vehicle trajectory patterns in dynamic settings \cite{vandenende}.

This study aims to demonstrate the effectiveness of DAS technology in monitoring urban traffic, particularly in densely populated areas with multi-lane roadways. We propose integrating DAS data with camera-based systems for a training stage, and relying exclusively on DAS data for continuous operation (Figure \ref{fig:main_diagram}). Our primary contributions are outlined as follows:
\begin{enumerate}
 \item Articulate the link between the camera and DAS model and outline the classification task (Section \ref{sub:camera_model}).
    \item Conduct the experiment described in \ref{sub:data_collection} and disseminate the results \cite{dataset}.
    \item Present a comprehensive pipeline for data labeling, integrating the camera and fiber systems \ref{sub:data_annotation}.
    \item Propose a novel regularization method to address challenges posed by noisy labeling \ref{sub:manifold_smoothness}.
    \item Define metrics \ref{sub:scores_metrics} to assess the performance of our model, along with an application example pertinent to smart cities \ref{sub:smart_city}.

\end{enumerate}

Section \ref{sec:background} briefly reviews the geophysical and computer vision background necessary for this study. In Section \ref{sec:method}, we present our approach, including the pre-processing stage and the neural network model. Section \ref{sec:experiment} covers the experimental results, including some limitations and post-processing improvements. Section \ref{sec:conclusion} is devoted to discussion and conclusions. Complementary material is provided in four appendices.

\section{Background}
\label{sec:background}

\subsection{The seismic wavefield generated by vehicles}
\label{subsec:siesmic_wavefield}
Moving vehicles generate a seismic wavefield that comprises of two major components \cite{vandenende, yuan2023, impulseresp1}. The first is the quasi static ($<$1 Hz) signal, resulting from subsurface deformation due to loading by the weight of the vehicle. The second is the dynamic component which is due to seismic waves, mostly surface waves, \cite{seismology-book1} generated by wheel-road interactions due to roughness of the road, especially bumps. These are generally observed in a frequency range of 2-20 Hz, but the exact values vary \cite{jousset2018dynamic, lindsey2020city, impulseresp1}. Previous studies \cite{vandenende}, \cite{impulseresp1}, \cite{impulseresp2} show that the Flamant-Boussinesq approximation adequately describes the instantaneous quasi-static deformation of the subsurface caused by an ideal point load at a given location. According to this approximation, the displacement along the $x$ direction, $u_x$, of a point $x, y, z$ in the subsurface, as a response to a point load at the origin, is:
\begin{equation} \label{eq:eq_u}
    u_x\left(x,y,z\right)=\ \frac{F}{4\pi G}\frac{x}{r^2}\left(\frac{z}{r}+\ \frac{2v-1}{1+\frac{z}{r}}\right) \ .
\end{equation}

The symbols in this equation are:
\begin{itemize}
    \item $x, y, z$ are spatial coordinates in 3-D space of the point at which the displacement is measured. We define the $x$ axis to be along the direction the fiber-optic cable, the $y$ axis complements it along the surface, and $z$ is the perpendicular depth beneath the surface.
    \item $r = \sqrt{x^2+y^2+z^2 }$ is the distance between the point load location and the location of the measurement.
    \item $F$ is the force applied by the point load on the ground.
    \item $G$ is the shear modulus of the ground, assumed to be homogeneous.
    \item $\nu$ is Poisson's ratio of the medium, assumed to be homogeneous.
\end{itemize}

\subsubsection{DAS Measurements}
\label{subsec:das_measurments}
In $\Phi-OTDR$, the measured phase changes are quasi-linearly related to strain or strain-rate along the direction of the fiber. The interferometric nature of the measurement is such that the phase difference, and thus strain, is computed over a subset of the array called a gauge length, and whose size is typically several to tens of meters \cite{vandenende, lindsey2021fiber}. Given a gauge length $L$, the quasi-static field induced by a vehicle and recorded by DAS will thus be the spatial derivative of the average displacement along the direction of the fiber (Eq. \ref{eq:eq_u}), computed over the gauge length L 
\begin{equation} \label{eq_e_prime}
    {{\varepsilon}}_{DAS}=\frac{1}{L}\left[{{u}}_x\left(x\ +\frac{L}{2},y,z\right)-\ {{u}}_x\left(x\ -\frac{L}{2},y,z\right)\right] \ .
\end{equation}

Furthermore, since the DAS system that we use measures the strain rate, an additional time-derivative is applied. The response of the fiber strain rate to a point load at the origin is thus:

\begin{equation} \label{eq:eq_e_dot}
    {\dot{\varepsilon}}_{DAS}=\frac{1}{L}\left[{\dot{u}}_x\left(x\ +\frac{L}{2},y,z\right)-\ {\dot{u}}_x\left(x\ -\frac{L}{2},y,z\right)\right] \ .
\end{equation}

We emphasize that vehicles are accurately described by a multi-point loading, applied at the points of contact of the wheels with the surface. The total recorded signal is thus a linear combination of several point loads, and multipole expansions of this equation are required \cite{yuan2023}.

In practice, setting up the IU involves setting specific temporal and spatial resolutions for sampling the fiber. Some DAS systems allow the control of the gauge length, but it is fixed in the IU we use. Acquired data are thus 2-D, with [time, distance along the fiber] dimensions. We refer to each spatial sampling point along the fiber as a channel. It is important to note that the channel spacing differs from the gauge length, and refers to the distance between successive measurement points. In this study, we use a channel spacing of 1m, and a gauge length of 10m.

\subsection{Camera model}
\label{sub:camera_model}
Let us define a general world coordinate system and an object $i$, such that the object's (vehicle) location is then represented by a three-dimensional vector $X_i$ related to the origin of the axis. Adopting the common pinhole camera model, we then can define the projection matrix $P$, which represents the mapping between the object three three-dimensional world location to its two-dimensional image location in pixels $x_i$ \cite{stockman2001computer}. Namely, for any object $i$: $ x_i = P X_i$. Mathematically, it is convenient to represent the fiber location as a 3D curved line $S$ and to parameterize it by the length parameter $l$, such that $S(l)$ is a 3D point along the fiber, with distance $l$ from its beginning. Similarly, the 2D camera projection of the fiber is $s(l) = P S(l)$.

The distance function between the fiber and object is given by:
\begin{gather} \label{eq:dist_eq}
    d\left(X_i, F(s)\right) = || X_i - S(l) ||^2 = || P^{-1} \left( x_i - s(l) \right) ||^2 \ .
\end{gather}
In general, the projection matrix is not invertible because it projects from 3D to 2D (any point $x_i$, $s(l)$ can be back-projected to a line instead of a point). However, we assume that the surface of the road can be represented as a 2D plane. In this case, the mapping between pixels and true location can be represented by an homography transformation \cite{szeliski2022computer}.

\subsubsection{Strain-rate map model}
\label{sub:strain_rate_model}

We introduce the concept of the strain-rate map as an extension of equation \ref{eq:eq_e_dot} as function of time $t$ and distance along the fiber $l$:
\begin{gather} \label{eq:phi_model}
    \Phi(t,l) = \int dV \dot{\rho}\left(\vec{r}, t, S, G, \nu, F, l \right)  \ ,
\end{gather}
With $\rho$ a phase-density distribution function. Namely, $\rho$ is the local contribution of each possible loading point, in space and time, to the phase shift measured at a certain distance along the fiber, $dV = dx\cdot dy\cdot dz$, and all other values are as previously defined. It is noteworthy that by approximating the scene as occurring on a 2D plane, we can redefine the volume integral as a surface integral. If the subsurface properties are approximately constant along the fiber, we can also neglect the spatial dependence of the density function on $G$ and $\nu$. In addition, we also assume that the fiber installation depth is constant and that the fiber is deployed in parallel to the road, simplifying that $x \approx l$. Whereas the last assumption is not necessarily correct for every point along the fiber trajectory, it is crucial because our study utilizes a spatially invariant model. Nevertheless, we later discuss a method to relax this assumption.

Under these assumptions, the density function becomes:
\begin{gather}
    \rho\left(\vec{r}, t, S, G, F, v \right) \approx \rho\left(d, t, F, l \right) \ ,
\end{gather}
With $d = d(\vec{r}, S(l))$ the distance between the loading and measurement points. We further assume a discrete number of loading points:
\begin{gather}
\label{eq:phase_shift_map}
    \rho\left(d, t, F, l \right) = 
    \sum_i \phi(d_i, c_i, t, l) \delta(\vec{d} - \vec{d_i}) \ .
\end{gather}
With $d_i, c_i, t, l$ representing the object distance from the fiber, class type, time, and the location along the fiber, respectively. By transitioning from the variable force $F$ to the class number $c_i$, we effectively categorize the objects into distinct equivalence classes, each identified by a class label. It should be noted that this approach is an approximation, as it treats different forces and thus vehicle weights within the same class as indistinguishable.

\begin{figure}
    \centering
    \includegraphics[width=1.0\linewidth]{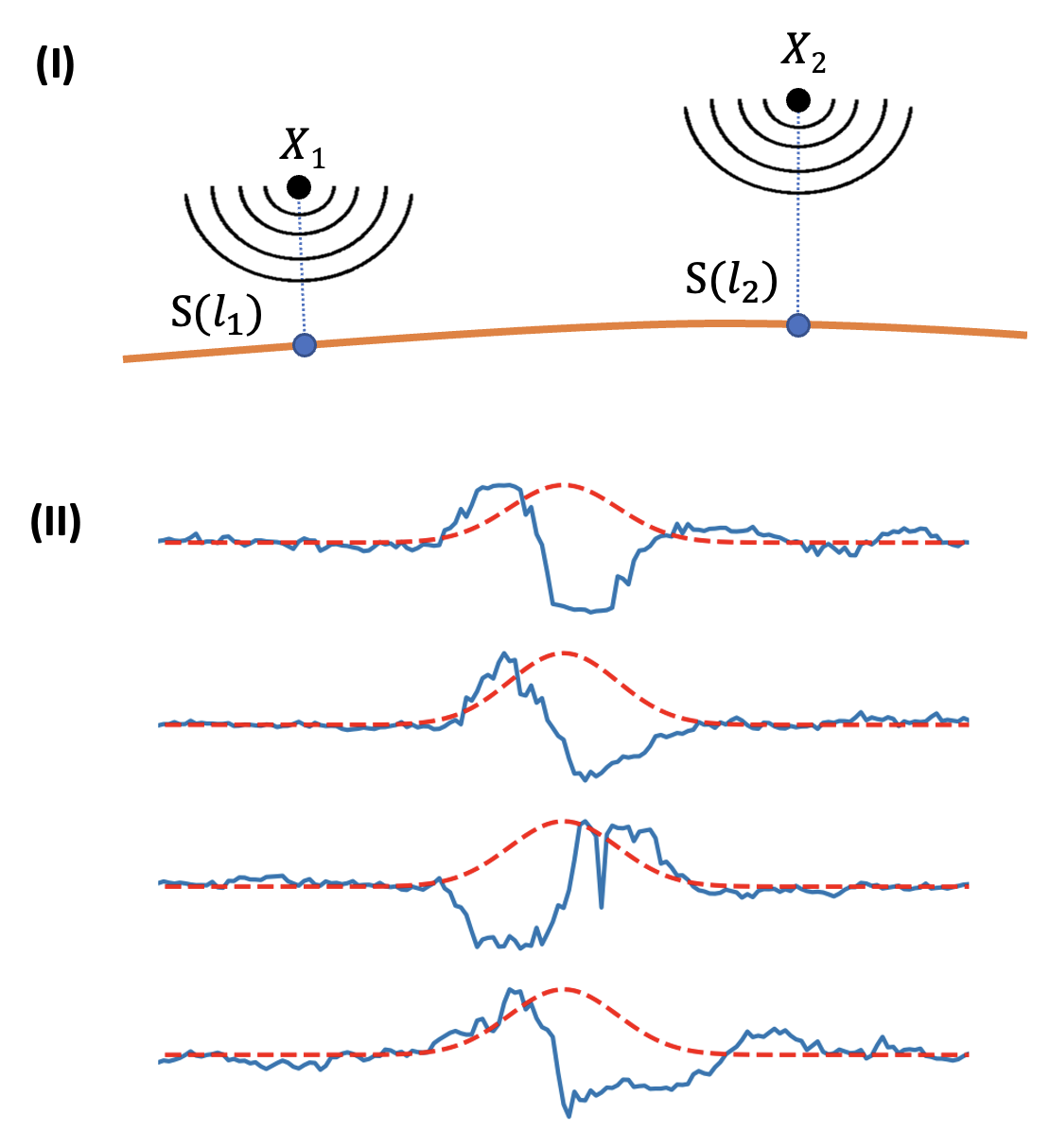}
    \caption{Strain rate in the Quasi-static approximation (constant time). (I) Each point along the fiber (orange) aggregates phase-shift contributions from all objects (denoted as $X_i$), $S(l_i)$ represent the point associated with this object along the fiber. (II) DAS recordings of the strain-rate (solid blue) and Gaussian probability distributions to detect an object at different locations (dashed red)}
    \label{fig:FOSinverse}
\end{figure}

Substituting into \ref{eq:phi_model}:
\begin{gather}
    \Phi(t,l) = \sum_i \dot{\phi}(d_i, c_i, t, l)  \ .
\end{gather}
Figure \ref{fig:FOSinverse} (I) illustrates the above model, in which the phase shift at time $t$ and distance along the fiber $l$ is the sum of the contributions from every loading point along the road. The contribution to the phase shift is determined by the location of each object and its weight, approximately represented by the object class, and can be expressed using the analytical expression introduced in \ref{subsec:siesmic_wavefield}.  However, we are interested in the probability distribution to detect an object at a specific location and time. Therefore, instead of the analytical solution, we associate a Gaussian probability distribution to the point along the fiber that is neareast to the object $S(l_i)$. Figure \ref{fig:FOSinverse} (II) shows the Gaussian probability distribution (red) associated with each point along the fiber $S(l_i)$, and its comparison with the real DAS recording data (blue). 

We choose not to use the analytical model directly as it contains many assumptions that may not be realistic in a real-life scenario - homogeneity of subsurface parameters, no influence of the precise car position along the road/lane, fiber installed parallel to the road, and accurate representation of a vehicle by a single point load. Whereas a full inversion approach may be theoretically possible, we opt for a simpler, more robust approach that does not depend on many unknown parameters.

\section{Method}
\label{sec:method}

\subsubsection{Detection map}
\label{sub:detection_map}
Similarly to \ref{eq:phi_model}, we define another function $D(t,l)$ for any set of object classes $C=\{c_1,c_2,..., c_N\}$ (e.g. cars or trucks).  This function specifies, at time $t$ and distance along the fiber $l$ , the probability that DAS sensed each of the objects in the class, including the zero object of noise only. We can formulate it as:
\begin{gather}
    D(t,s) = 
    \begin{bmatrix}
        \text{P} {(c=c_1)} \\
        \vdots \\
        \text{P}(c=c_N) \\
        \text{P}(c\notin C) \\
    \end{bmatrix} \ .
\end{gather}

\subsubsection{Neural Network}
\label{sub:neural_network}
We use a deep learning model $\mathbf{T}$ to learn the map between the phase-shift function and the detection map:
\begin{gather}
    \hat{D}(l,t) = \mathbf{T}\left[\Phi(l,t)\right] \ .
\end{gather}
Whereas this mapping can be conducted in various ways, we choose a UNET architecture \cite{Unet, pix2pix2017} to leverage the high spatio-temporal coherency offered by DAS. The training process follows a self-supervised regime in which the processing of visual data from the camera yields the ground truth detection map. We use the following loss function:
\begin{gather}
    \mathcal{L} = - \mathrm{E} \left[ \sum_{c} D(l, t, c) \log \hat{D}(l, t, c)   \right] \ .
\end{gather}

\subsubsection{DAS Pre-processing}
\label{sub:pre_processing}
To speed up the convergence of the network training, we apply standard preprocessing to the DAS data to highlight physically meaningful signals. First, we apply a sample-by-sample and channel-by-channel median filter, which mitigates instrument noise and optical response variations along the fiber. Then, we apply low-pass filtering and downsampling to 30 Hz to match the visual sampling frequency and minimize computation time. Finally, we apply 2-D FK-filtering to maintain signals with phase velocities between $2-90$ km/h and frequencies below 1 Hz, as this is the range of quasi-static signals excited by moving vehicles. We provide additional technical details on the preprocessing stage in Appendix \ref{app:preprocessing}

\section{Experiment}
\label{sec:experiment}

\subsection{Data collection}
\label{sub:data_collection}

The experiment was conducted for about a week during March 2023, at Klausner Street near the Tel-Aviv University in Israel \footnote{\href{https://maps.app.goo.gl/GknQ7Rcbbv1mXrhn7}{Google maps}}. There is ample traffic to and from the university and a nearby science museum, including multiple regularly operating bus lines.  Whereas DAS recording took place for a week, we recorded 9 hours (8AM - 5PM) of a weekday with a smartphone camera (Samsung Galaxy 23 Ultra) mounted on a tripod on top of a nearby university building, yielding a diagonal point-of-view (See Figure \ref{fig:DAS_vs_Camera}). The portion of the fiber along the road is approximately 350 meters long, with the camera covering approximately 80 meters of the road. In this study, we train the model and assess its performance using the section observed by the camera as we have ground truth labels, but apply the trained DAS network to the entire fiber. The dataset is accessible in \cite{dataset}.

\subsection{Spatio-temporal Calibration}
\label{sub:spatio_temporal_calibration}
While we had rough maps outlining the installed fiber's path, we conducted a manual calibration process to accurately align the video recordings with the DAS measurements.
This step is crucial as we assume perfect spatio-temporal synchronization between the fiber and the camera. For the calibration, we selected 80 reference anchors along the anticipated fiber path covered by the camera. Then, we struck the ground at these anchors with a hammer, recording the exact times and channels along the fiber where the impacts were most pronounced. To locate the fiber in the 3D world coordinate system, we used the projection matrix (as explained in \ref{sub:camera_model}) and assumed, based on infrastructure maps, that the fiber is buried at 90 cm depth. Our findings suggest that the fiber runs primarily parallel to the road and deviates by at most 1 m in the perpendicular direction.  Throughout the experiment, we assume that the projection matrix remains constant thanks to the stability of the camera. Figure \ref{fig:DAS_vs_Camera} schematically marks the fiber's position along the street.

\subsection{Data Annotation}
\label{sub:data_annotation}
\begin{figure}
    \centering
    \includegraphics[width=1.0\linewidth]{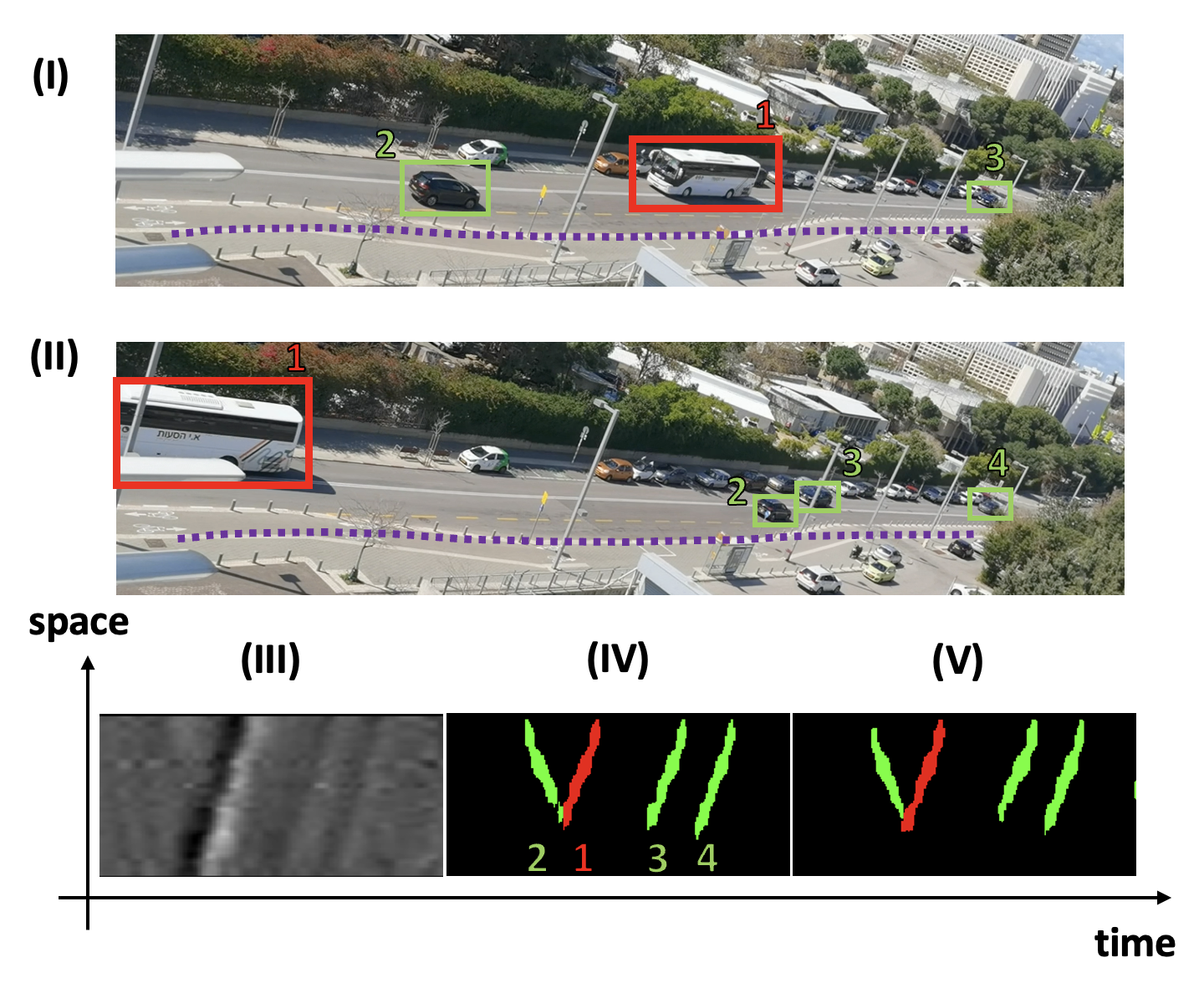}
    \caption{Comparison between the camera and the DAS . (I) First frame of the scene, with 3 labels: 2 cars and a single bus. The purple line represents the estimated location of the fiber (based on the calibration process). (II) Last frame of the scene, with 4 labels: 3 cars and a bus. (III) Recorded DAS data after pre-processing. (IV) Generated labels from camera (V) The output of the model based on DAS only.}
    \label{fig:DAS_vs_Camera}
\end{figure}
To generate ground truth labels from the visual data, we use the YOLOv11 Ultralytics\cite{yolov8Ultralytics} algorithm on the recorded video. This algorithm detects, tracks, and classifies objects in the scene such as cars, trucks, buses, pedestrians, motorcycles, and more. In this study, we focus on three classes: cars (small vehicles), buses or trucks (large vehicles), and noise. We have also attempted to expand our model classes to trucks only and motorcycles. However, the results were unsatisfactory, mainly due to the uneven class distribution in the data (many more buses than trucks) and the poor DAS signal-to-noise ratio of motorcycles. After the detection, we filter out static objects, such as parked cars, as they do not generate any strain-rate signal. 

Then, to annotate any object in the detection map, we use the Gaussian envelope of the phase-shift function, as depicted in Figure \ref{fig:FOSinverse} (II). Practically, for each moving detected object $X_i$, we find the location among the fiber $l^*$ which minimizes the distance (see equation \ref{eq:dist_eq}):
\begin{gather}
    l^*=\text{argmin}_{l} [d\left(X_i, S(l)\right)] \ .
\end{gather}
This defines the detection line along the DAS image. Then, we use a Gaussian filter to smear the probability of detection along time and space, represented by DAS channels. The Gaussian's standard deviation was chosen to be around 1/12 seconds on the temporal axis and approximately 2.5 meters on the spatial axis, with both equating to 2-3 pixels. These parameters were chosen empirically and were highly correlated with the functional strain-rate shape, as can be seen in Figure \ref{fig:FOSinverse} (II). Visually, it can be seen that they generally encapsulate the high-energy portion of the signal, but we again emphasize the approximate nature of this formulation. 

Thus, we obtain a detection map $D(t,l)$, as shown in Figure \ref{fig:DAS_vs_Camera}, which expresses the probability of detecting each of the objects. The probability is normalized between 0 to 1 and sums up to 1 among the channels, namely $|| D_c(t,s) ||_1 = 1$. The normalization is provided by the softmax activation function utilized in the UNet architecture's output.

\subsection{Training the Neural Network}
\label{sub:training_nn}
The data were shuffled and divided into batches consisting 48 samples each, such that any sample covers approximately about 30 seconds. 
Next, we partitioned $70\%$ of the data for training purposes, $15\%$ for validation, and $15\%$ to serve as a test set for which we report final results. 
We perform augmentation on the training dataset by applying both horizontal and vertical flips. Horizontal flipping acts as a time reversal, whereas vertical flipping represents a spatial reversal, similar to flipping from north to south.
This augmentation was done in order to minimize inherent biases in the data due to uneven lane usage and stronger signals for southbound traffic due to proximity to the fiber.
The training process lasted 100 epochs, employing a learning rate of 0.001 and a weights decay coefficient of 0.001, incorporated inside the AdamW optimizing algorithm \cite{adamw}. This algorithm combines the advantages of adaptive weight updates with regularization techniques to encourage better generalization and prevents over-fitting. Using AdamW, our aim is to strike a balance between efficiently updating the network weights based on the gradients and regularizing the model to improve its overall performance. To optimize the network, we employ the Cross Entropy Loss function, which allows us to effectively measure the dissimilarity between the predicted probabilities and the ground truth labels. We also tried to optimize the model using Dice loss, L1, L2, and Huber loss, but all yielded subpar results.

\subsection{Score metrics}
\label{sub:scores_metrics}
To give an objective and independent metric to score our performance, we separate evaluation metrics for detection and classification, as derived in Appendix \ref{app:metrics_dev}. For any pair of detection maps $D$ (labeled), and $\hat{D}$ (model output), we define:
\begin{align}
    C_{\textbf{DET}} =&   P\left[ \bigcup_{c\in \{\text{n}, \bar{\text{n}}\}} \left( \hat{D} = c\ \bigcap D = c \right) \right]  \ , \\
    C_{\textbf{CSF}} = & \frac{P\left[ \bigcup_c \left( \hat{D}=c \bigcap D=c \right) \right] }
    {  P\left(  D \neq \text{Noise} \right) } \ ,
\end{align}

While $\text{n}$ stands for noise, and $\bar{\text{n}}$ for no-noise, namely, the complementary set. 
Additionally, we evaluated the Dice loss and the False alarm \cite{dicelossref}. The Dice loss was proven effective for alignment tasks because it emphasizes the overlap between predicted regions and actual regions of interest. 
Table \ref{tab:summery_of_results} shows a summary of our results, indicating the validity of the approach.

\begin{table}
    \centering
    \begin{tabular}{|c|c|c|c|c|} \hline 
         Detection $\uparrow$ &  Classification $\uparrow$ & Dice Loss $\downarrow$ & False Alarm $\downarrow$ \\ \hline 
         94.2\%&  94.0\%& 0.581& 1.28\%\\ \hline
    \end{tabular}
    \caption{Results summary of the model. The arrows indicate the direction of the better results.}
    \label{tab:summery_of_results}
\end{table}

\subsection{Case-studies and Noisy Labeling}
\label{sub:case_study_noisy_label}
Utilizing visual-based algorithms such as YOLO for data labeling often introduces some noise. For example, some vehicles may be missed or improperly segmented during detection. Here, we analyze both the successes and failures of the model, which we find to be strongly correlated with the noisy labeling. 

\begin{figure}
    \centering
    \includegraphics[width=1.0\linewidth]{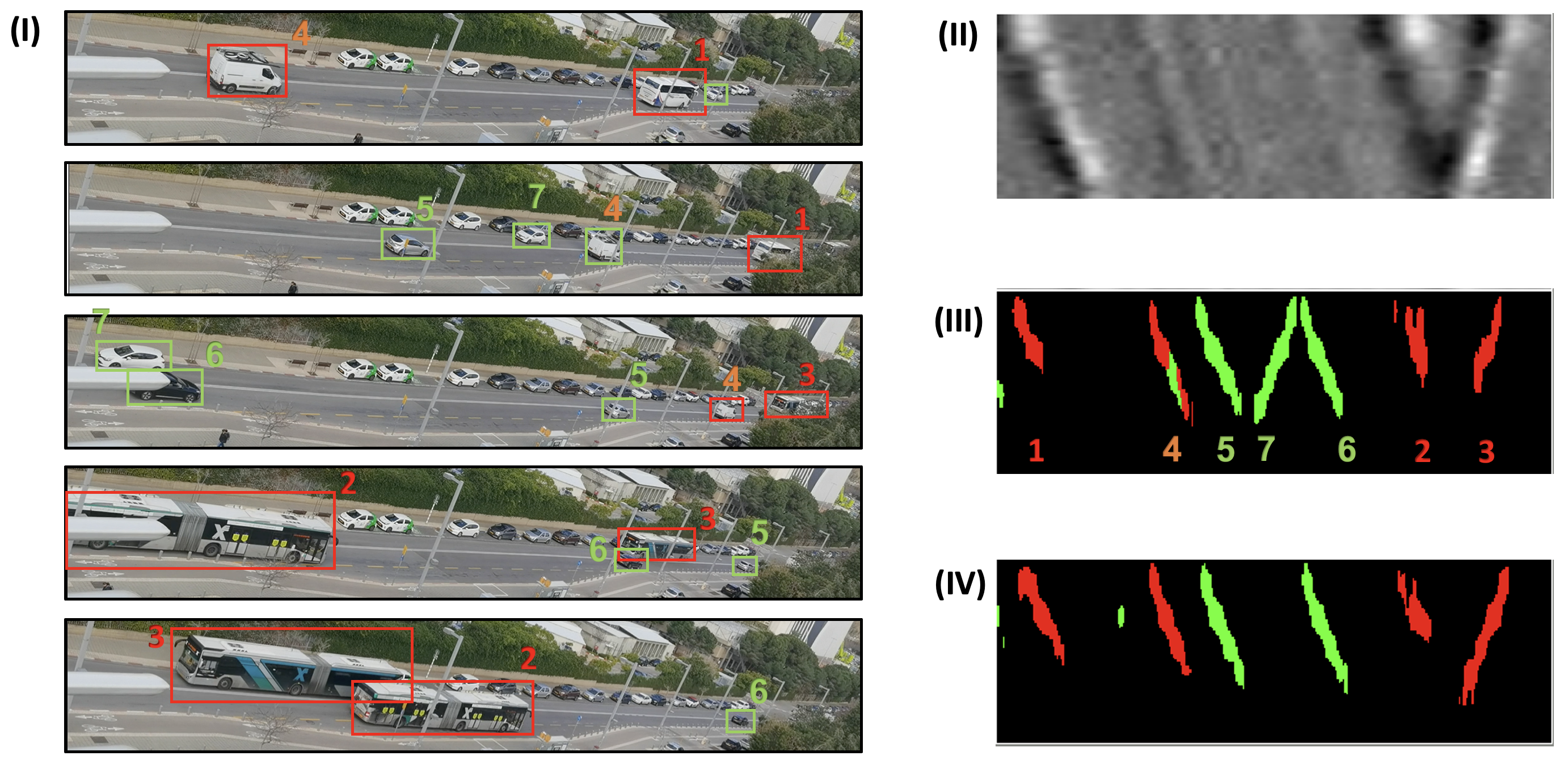}
    \caption{Use-case analysis of the model's performance. (I) 5-frame sequence and itslabeling with rectangular boxes. (II) The DAS recording in this time interval (model input). (III) Labeling from visual input (YOLO). (IV) DAS model output results. We consider vehicles 1-6 as success cases, and vehicle 7 as a failure case.}
    \label{fig:Sequence_example}
\end{figure}

Figure \ref{fig:Sequence_example} provides a detailed example showcasing results with 7 vehicles observed in a short time span of approximately 30 seconds. Most vehicles were correctly detected and classified by the model, butvehicle number 7, moving in the father lane from north to south, was entirely overlooked. An interesting event occurs with vehicle number 4, where the visual-based algorithm (YOLO) altered its classification across frames, but ourmodel classified it as a constant class, indicating some ability of our network to deal with errors in the labeling. Another notable success is observed in point 3, where the DAS model effectively identified the bus over a greater distance than the vision-based algorithm (YOLO).
It is also apparent that there is an additional green patch located between 1 and 4 in the model's output, likely due to noise in the signal. Although this counts as an error in the loss calculation, it will not be regarded as an object since its area is too insignificant (for example, see our counting algorithm in \ref{sub:smart_city}).

Empirical evidence shows that the model primarily underperforms when the SNR is very low, typically at thefarther lane from the fiber. Additionally, the training labels are often quite noisy, yet interestingly, the neural network occasionally manages to mitigate their negative effects and accurately detect and classify vehicles.

\subsection{Manifold smoothness - Dealing with noisy labeling}
\label{sub:manifold_smoothness}
To overcome noisy labeling consequences, we suggest adding a prior to the model's output. We use a masked version of the total variation loss, working only along the 1-dimensional trajectory (1D manifold). The underlying idea is to reduce the gradients across each detection line, thereby motivating the model to favor "smooth" configurations.:
\begin{gather}
    \mathcal{L}_{Smooth} = || \nabla_{M} \hat{D}(l,t) ||^2 \ ,
\end{gather}
While $\nabla_M \equiv M(x,y) \cdot \nabla$. Namely, for a given mask $M$, where $M=1$ inside an arbitrary domain (1D trajectories in our case), and $M=0$ anywhere else, $\nabla_M$ calculates the directional gradient in that domain and $\nabla_M =0$ anywhere else. To calculate $M$ we used the well known Hough transform algorithm \cite{houghTransform}, with further details given in Appendix \ref{app:manifold_smooth}

\begin{figure}
    \centering
    \includegraphics[width=1.0\linewidth]{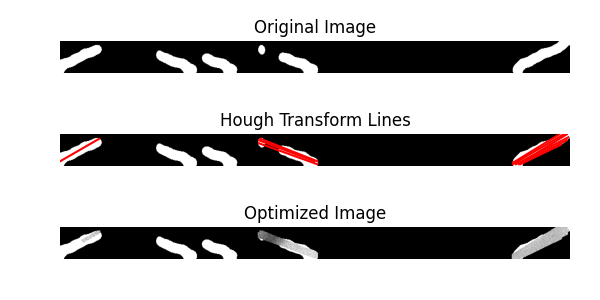}
    \caption{The detection map of a single class. A successful example of the smoothing post-processing stage. However, as seen, not all the lines were detected.}
    \label{fig:HG1}
\end{figure}

This loss function can be used as a regularizer during the training, or as a post-processing fine-tuning stage. In our tests, we limit ourselves to the case of post-processing stage: after the training process, we performed another fine-tuning training to minimize the $\mathcal{L}_{Smooth}$ term. Qualitatively, we found that this approach leads to more reasonable results (e.g. it transforms any dashed or segmented line into a solid one), as can be seen in Figure \ref{fig:HG1}, but that is also requires intensive manual calibration in tuning the Hough transform parameters. We leave room for improvement in this method to future studies.

\section{Discussion}
\label{sec:app_and_discussio}

\subsection{Performance and Methodology}

Our results demonstrate high accuracy for detection and classification, exceeding 94\%, along with a low false alarm rate (about 1.5\%). Such performance was achieved through a data collection strategy that involved recording for several hours and extrapolating to a whole week. Spatially, we recorded 80 meters of the road and extrapolated for the remaining 270 meters. This approach allowed us to generate comprehensive statistics and analysis on vehicle counting, classification, and speed estimation.

\subsection{From Sensing to Smart City applications}
\label{sub:smart_city}
\begin{figure*}
    \centering
    \includegraphics[width=1.0\linewidth]{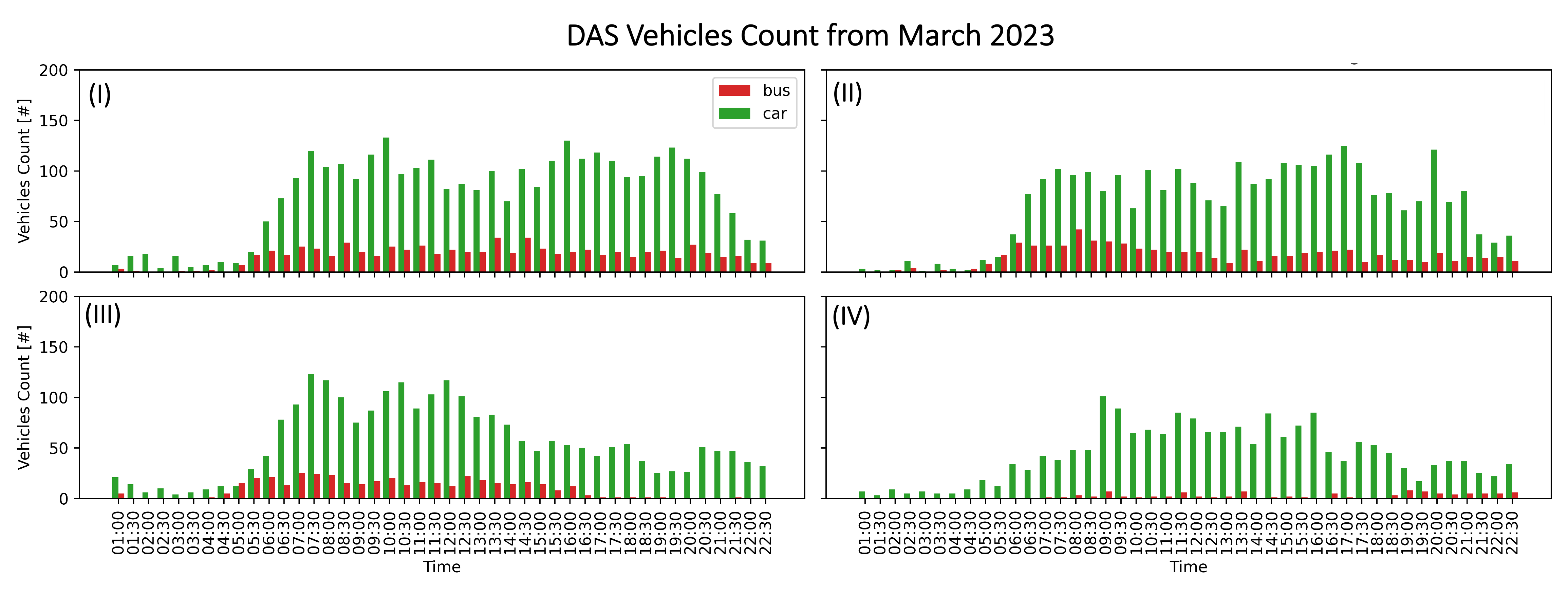}
     \caption{Collecting statistics for smart city applications - vehicles counting throughout the week. (I), (II), (III), and (IV) represents Wednesday, Thursday, Friday and Saturday, respectively.}
     \label{fig:statistics_along_week}
\end{figure*}

To assess our model's performance and its advantages, we analyze vehicle statistics throughout the week. Specifically, we apply the trained model to unseen DAS recordings to obtain distributed statistics on vehicle traffic along the fiber. Figure \ref{fig:statistics_along_week} presents the vehicle counts for four different days of the week for the whole fiber distance (about 350 meters). 

The analysis uncovers a variety of intriguing trends, including the usual start time of the workday, the average vehicle density, and the ratio of buses to cars. This emphasizes the street's heavy usage by buses, consistent with at least ten different regularly operating bus lines along this route. 
Saturday, the rest day in Israel, exhibits an entirely different pattern: a slow rise in activity, fewer vehicles, and almost no buses\footnote{In Israel, public bus services predominantly cease operations on Saturdays, adhering to cultural and religious practices}. Friday afternoon, which is approximately the time at which bus service stops, also shows a very clear reduction in bus usage.
We show more statistical data about phase distribution in Appendix \ref{app:phase_statistics}  and about the vehicles' speed in Appendix \ref{app:speed_statistics}.

\subsection{Assumptions and Limitations}
\label{sub:discussion_and_limitations}

\subsubsection{Data Quality and Model Robustness}

A significant challenge encountered was the presence of noise in the labels. Interestingly, the model occasionally demonstrated an inherent ability to reduce noise effects. To further mitigate this source of noise, we proposed a method to smooth the detection map. Despite the high performance of the model, we believe that improving the labeling quality will dramatically improve its performance and generalization abilities.

\subsubsection{Strain-Rate Shape and Gaussian Envelope Analysis}

We examined the relationship between the strain-rate shape and the Gaussian envelope, finding similarities but not consistent equivalence. Comparisons with analytical functions revealed discrepancies, arising from deviations from our initial assumptions, as described earlier. We addressed these limitations through empirical tuning of the Gaussian envelope, which proved effective in practical applications. It possible that with larger, properly labeled datasets, more complete signal representations that further resemble the expected analytical behavior can be utilized and leveraged for more accurate results. 

\subsubsection{Transferability and Scalability}

Our approach demonstrates promising transferability to new sites. While we used 9 recording hours for optimal training, we found that only a few hours of data collection could yield at least 90\% of the optimal performance given nominal traffic conditions. Similar scalability was observed in the spatial domain, where training on half of the visually-covered fiber (40 m instead of 80m) was sufficient for accurate predictions (at least 90\% of the optimal performance) over larger distances (the whole 350 meters). These results suggest that our approach could be readily adapted to new fiber layouts with minimal site-specific data collection. 

\subsubsection{Model Assumptions and Real-World Applications}

A key assumption in our model is the parallel alignment of the fiber to the road, which may not always be true in real-world scenarios. This assumption allowed us to employ a spatially invariant model, such as a fully Convolutional Neural Network. For cases where the fiber is not parallel to the road, future work could explore spatial encoding techniques or non-spatially-invariant architectures to extend the model's ability to analyze differently the input for different locations. It is also important to note that in practical deployments, the presence of fiber spools is very common, and may introduce complications. However, these can be readily detected and excluded from the DAS map, either through tap testing or other channel mapping methods, thus minimizing their impact on overall performance.

\subsubsection{Classification Limitations}

While the presented method supports a general number of classes, this study focused primarily on two vehicle types: cars and heavy vehicles (trucks and buses). This choice was primarily due to data imbalance issues, which posed significant challenges in detection and segmentation processes. The street under study had a high frequency of cars and buses but fewer trucks, motorcycles, and pedestrians. In addition, the weights of buses may significantly vary based on the number of passengers and whether the bus is articulated or not. Furthermore, the low signal-to-noise ratio (SNR) in detecting lighter objects, such as pedestrians, presents a challenge that may require more sensitive interrogators and specialized processing techniques beyond the scope of this project.

\section{Conclusion}
\label{sec:conclusion}
We present a hybrid approach, leveraging both visual and DAS data, for traffic monitoring in an urban environment. Our methodology enables precise detection, classification, and counting through neural networks applied to DAS data, with a training stage guided by video recordings. Our approach highlights the potential of combining fiber-optic sensors with visual cues, focusing on practicality and scalability, protecting privacy, and minimizing infrastructure expenses, making it a feasible choice for real-world traffic monitoring.

\section*{Acknowledgment}
We would like to thank Oz Matoki, Roy Mazuz, and Elyasaf Lichtenstadt for their help in the experiment and the preprocessing stages. The research was supported by the Shlomo Shmeltzer Institute for Smart Transportation at Tel-Aviv University.

\printbibliography

\appendix
\subsection{Pre-processing}
\label{app:preprocessing}

In the preprocessing stage, raw data obtained from fiber optic sensors undergo several essential transformations to enhance its quality and extract meaningful information:
\begin{enumerate}
\item Multi-channel Median Filter - To reduce background noise, we use a median filter for each channel separately, namely, for each location along the fiber. This step increases the SNR of the DAS signal.
\item Temporal Low-pass Filtering and Downsampling - The raw data is recorded in very high temporal frequency ($1$Khz), while for our purposes we don't need such high temporal resolution. Therefore, and to further attenuate high-frequency noise components, the data undergoes temporal low-pass filtering. Then, we downsample the data to match the same temporal resolution of the camera (In our case, $30$Hz). This matching is not a must but we found it helpful in the calibration process and the algorithm.
\item FK Filter - The FK-filter is a spatio-temporal frequency filter which is necessary to filter-out the non-relevant spectral contents. This filtering technique enables the extraction the data which is not in the interval between velocity interval, typically for traffic is in the range of $2$ km/h to $90$ km/h.
\end{enumerate}

\begin{figure}
    \centering
    \includegraphics[width=1.0\linewidth]{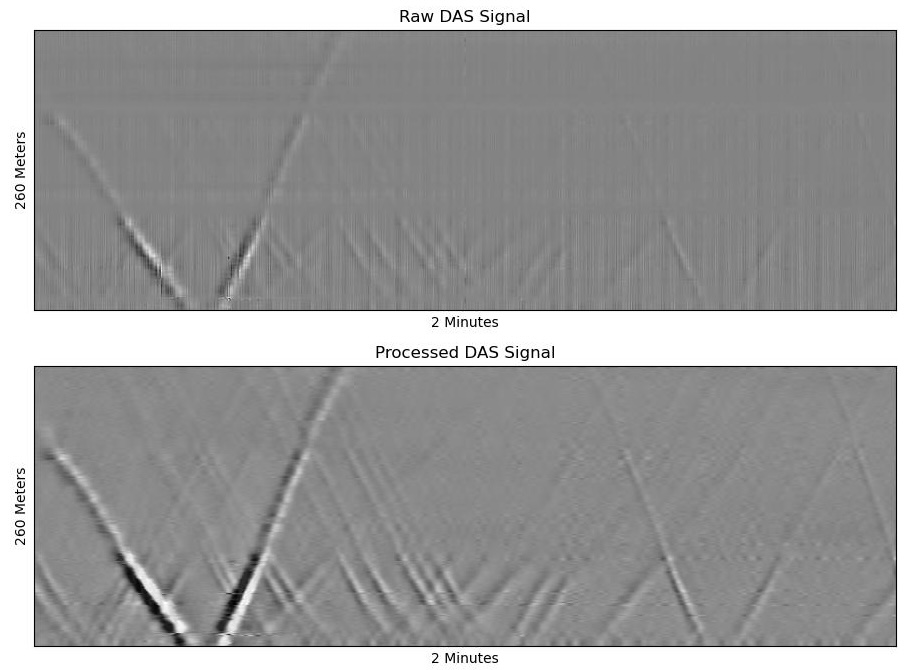}
     \caption{2 Minutes of DAS Data Before and After Processing}
     \label{fig:das_preprocess}
\end{figure}

A visual example is shown in \ref{fig:das_preprocess}

\subsection{Metrics derivation}
\label{app:metrics_dev}
For each pair $D$ (label) and $\hat{D}$ (model's output) we define the following metrics:

\subsubsection{Detection} -
We cosider a detection of an object, as the probability to distinguish it well from noise. Therefore, we define the score as follows:
\begin{gather}
        C_{\textbf{Detection}} =  P\left[ \bigcup_{c\in \{\text{n}, \bar{\text{n}}\}} \left( \hat{D} = c\ \bigcap D = c \right) \right] \\
\end{gather}
While $\text{n}$ stands for noise, and $\bar{\text{n}}$ for no-noise, namely, the complementary set. Note that this score is claculated pixel-wise, therefore, we also calculate the average over all the pixels. Explicitly:

\begin{gather}
        C_{\textbf{Detection}} = \frac{ 1 }{LT} \sum_p \bigg[  P \left( \hat{D}_p = \text{n} \right) P \left( D_p =  \text{n} \right) \\ \nonumber
        + P \left( \hat{D}_p = \bar{\text{n}} \right) P \left( D_p =  \bar{\text{n}} \right) \bigg]
\end{gather}
While $L$ and $T$ are the number of pixels in the space and time axes, respectivly. It is important to note that guessing a consant value, namely $P(\hat{D}_p=\text{n})=\alpha$ leads to a guessing score of:
\begin{gather}
        C_{\textbf{Detection}} =\frac{ 1 }{LT} \sum_p  \bigg[  \alpha  P \left( D_p =  \text{n} \right) \\ \nonumber + (1-\alpha) P \left( D_p =  \bar{\text{n}} \right) \bigg]
\end{gather}
Assuming that for any pixel $p$, the probability to get noise is higher than non-noise: $P \left( D_p = \text{n} \right) > P \left( D_p =  \bar{\text{n}} \right)$ and we get a maximum detection score by guessing $\alpha=1$, which leads to $ C_{\textbf{Detection}}= \text{Num of noisy pixels}/\text{Total number of pixels}$. Practically, this number varies between 0.5 to 0.8.

\subsubsection{Classification} -
We consider a classification score similar to the detection score, with a non-noise normalization factor:
\begin{gather}
    C_{\textbf{Classification}} = \frac{P\left[ \bigcup_c \left( \hat{D}=c \bigcap D=c \right) \right] }
    {  P\left(  D \neq \text{Noise} \right) }
\end{gather}
Similarly, this score is done for any pixel in the image

\subsubsection{Dice Loss}-
The Dice loss is defined as follows:
\begin{equation}
\mathcal{L}_{\text{Dice}} = 1 - \frac{2 |A \cap B|}{|A| + |B|}
\end{equation}
where A and B are two sets, defined for $D$ and $\hat{D}$, respectivly and for each class (except from noise). The sets are defind by the pixels that have higher probability distribution for this channel, indicating a class was detected. The threshold was tuned manually to be 0.5.

\subsection{Manifold smoothness}
\label{app:manifold_smooth}

\begin{gather}
    \mathcal{L}_{Smooth} = || \nabla_{M} \hat{D}(l,t) ||^2
\end{gather}
While $\nabla_M \equiv M(x,y) \cdot \nabla$. Namely, for a given mask $M$, where $M=1$ inside an arbitrary domain (1D trajectories in our case), and $M=0$ anywhere else, $\nabla_M$ calculates the gradient in that domain and $\nabla_M =0$ anywhere else. To calculate $M$ we used the well known Hough transform algorithm

\textbf{The process goes as follows:}

For any detection map, we can create a threshold image of it (convert it to a binary image of 0 or 1, according to the detection probability for each channel).
Then, on the thresholded image, we apply Hough Transform over the image. to get a set of detected lines: $\{ (x_1, y_1, x_2,y_2)_l \}_{l=1}^n$ (denote $n$ as the number of detected lines in the image).

For each line, we run from its starting point $(x_1,y_1)$ to its final point $(x_2,y_2)$ with a loop parametrized by a $t$ parameter, and the directional gradient of the detection map $D(x,y)$ is then calculated on each point along the line: $\nabla_M \equiv \hat{t} \cdot (\partial_x, \partial_y)$ while $\hat{t}$ represents the unity vector along the line.

The $\mathcal{L}_{Smooth}$ is given by accumulating the absolute value of all the directional gradients for each of the $n$ lines, and averaging over all the lines in the image.

Minimizing $\mathcal{L}_{Smooth}$ over the pixels domain, lead the detection map to be more smooth (in the directional gradient sense). We trained the detection map in an unsupervised way for 50 epochs with Adam optimizer and learning rate of 0.1. 

\begin{figure}
    \centering
    \includegraphics[width=1.0\linewidth]{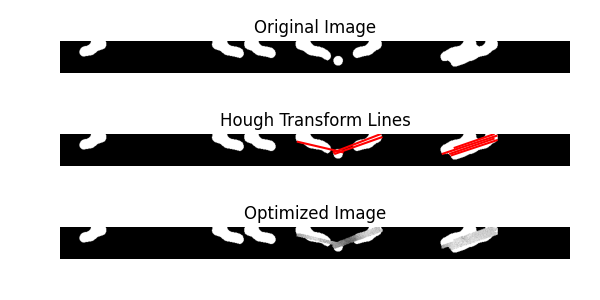}
    \caption{Example of the smoothing post-processing stage.}
    \label{fig:HG2}
\end{figure}

An example of the results can be shown in Figures \ref{fig:HG1} , and\ref{fig:HG2}.

\subsection{Phase Pattern Statistics}
\label{app:phase_statistics}

\begin{figure}
    \centering
    \includegraphics[width=1\linewidth]{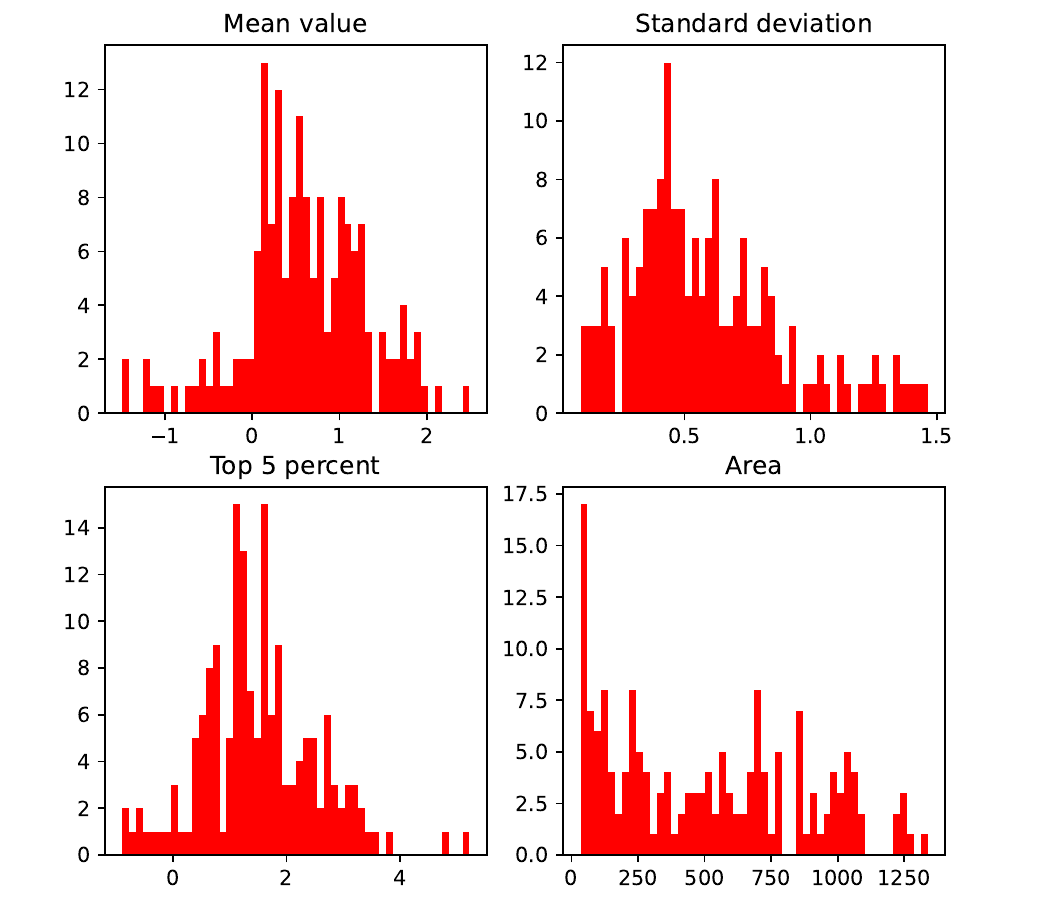}
    \caption{Four statistical measurement over all the buses for an entire day, on Wednesday, March 22, 2023.}
    \label{fig:stat_buses}
\end{figure}

\begin{figure}
    \centering
    \includegraphics[width=1\linewidth]{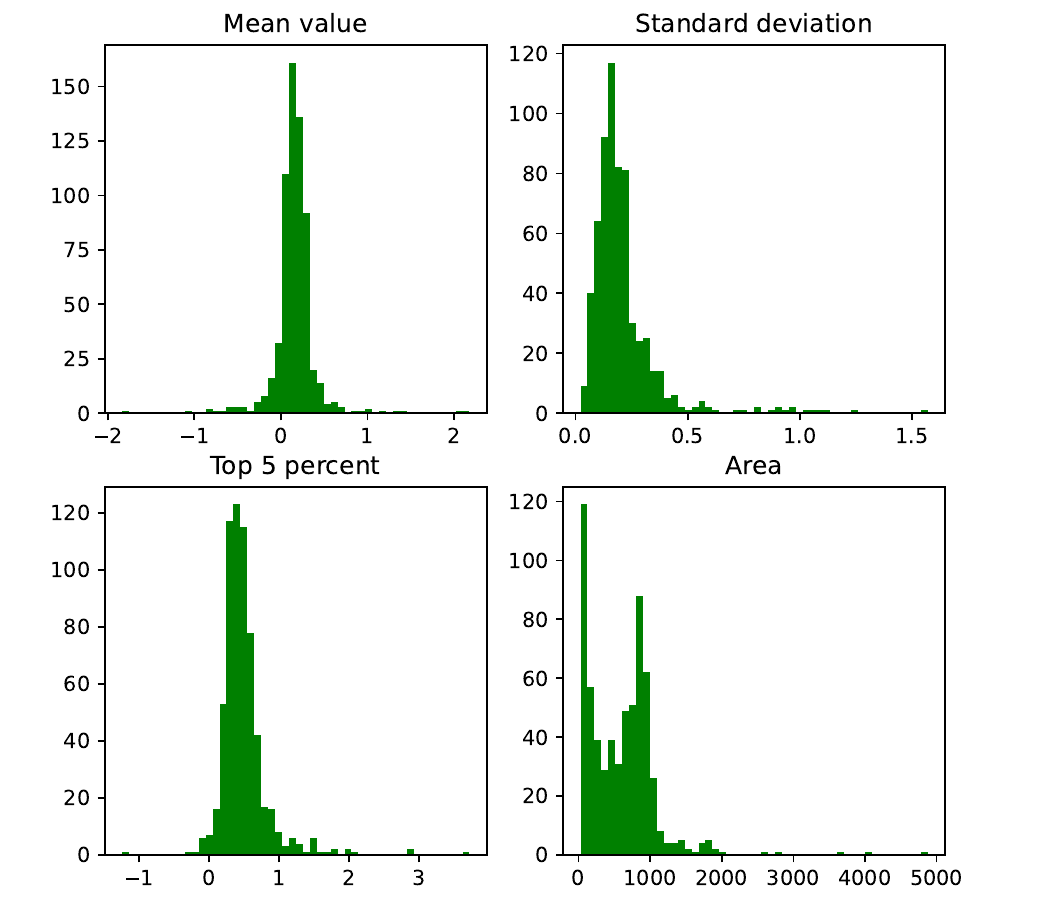}
    \caption{Four statistical measurement over all the cars for an entire day, on Wednesday, March 22, 2023.}
    \label{fig:stat_cars}
\end{figure}

Figures \ref{fig:stat_cars} and \ref{fig:stat_buses} present statistical information about cars and buses, respectively, for an entire day. Each statistical measure is derived from the phase shift recorded by the DAS:

\begin{enumerate}
    \item The mean value represents the average phase shift.
    \item The standard deviation quantifies the variation in the phase shift distribution.
    \item The area denotes the count of pixels identified as an object.
    \item The top 5 percent indicates the value at the $5\%$ upper end of the distribution for each object.
\end{enumerate}


\subsection{Speed Statistics}
\label{app:speed_statistics}

\begin{figure}
    \centering
    \includegraphics[width=1\linewidth]{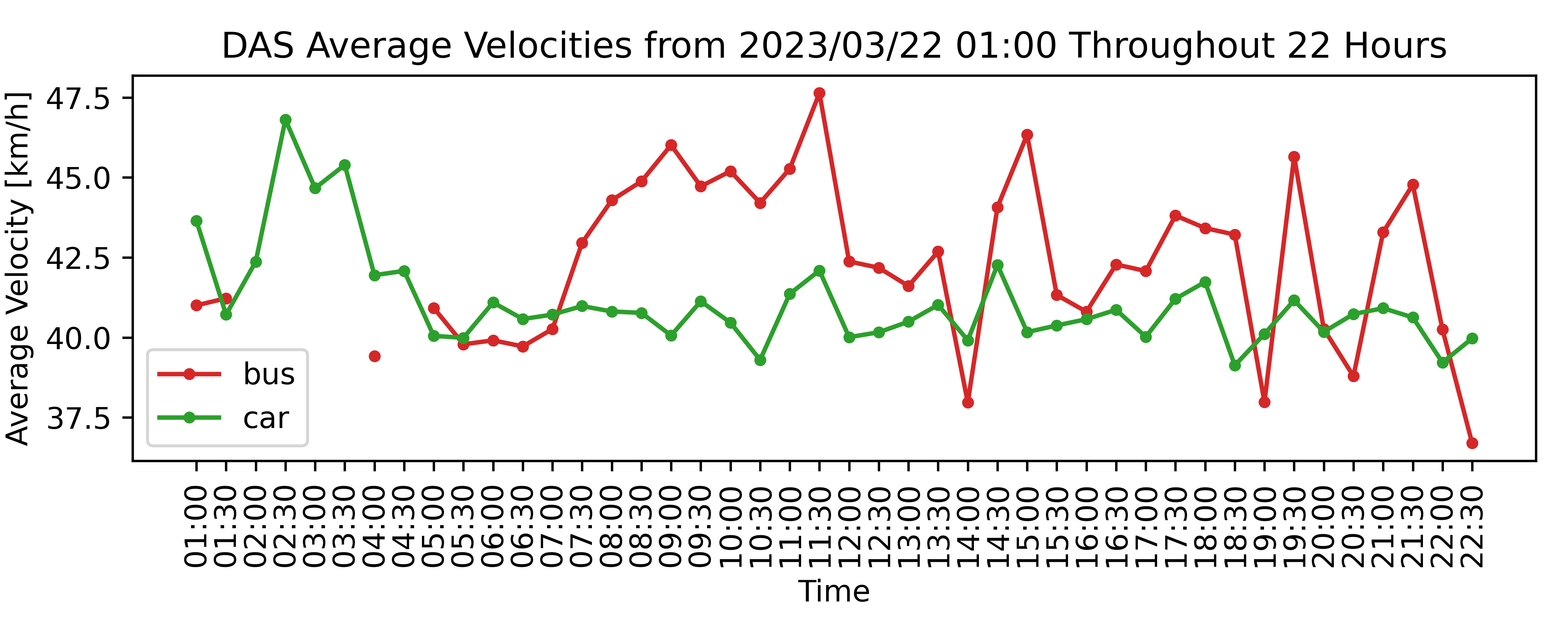}
    \caption{The speed of vehicles on Wednesday, March 22, 2023, throughout the day on the road.}
    \label{fig:velocity_stats}
\end{figure}

Figure \ref{fig:velocity_stats} shows an example of the speed distribution along a single day.

Building on the prior understanding that contours identified as vehicular objects often resemble tilted ellipses, we heuristically estimate the tilt angle by analyzing the standard deviations of pixel coordinates along both axes, $\sigma_x$ and $\sigma_y$. The angle $\theta$ is calculated as:

\begin{equation}\label{eq:angle-extraction}
\theta =\arctan\left(\frac{\sigma _y}{\sigma _x}\right)
\end{equation}
It is important to note that Eq. \ref{eq:angle-extraction} does not account for the direction of travel, as our primary goal is to compute the speed. If distinguishing the driving direction is required, the tilt orientation can be determined from the pixel coordinates, with an angular adjustment of $90^{\circ}$.

The computed angle is then converted to velocity by applying units conversions that depend on the sampling rates in both the temporal and spatial axes, as well as reversing any resizing operations performed during the learning pipeline.

\end{document}